\documentclass[12pt]{iopart}

\usepackage{bm}
\usepackage[english]{babel}
\usepackage[dvips]{graphicx}
\usepackage{latexsym}
\usepackage{amsthm}
\usepackage{amsfonts}
\usepackage{amssymb}

\usepackage[colorlinks=true,citecolor=blue,linkcolor=red,urlcolor=blue, linktocpage=true,breaklinks=true,pagebackref=false]{hyperref}

\usepackage{textcomp} 		
\usepackage{hyperref} 		
\usepackage{ifthen} 			
\usepackage{xifthen} 		
\usepackage{color} 			
\usepackage{soul} 			
\usepackage{lineno}
\usepackage[normalem]{ulem}
\usepackage{siunitx}
\usepackage{xcolor} 
\usepackage{float}

\usepackage{bbold}
\usepackage{mathrsfs}

\usepackage[latin1]{inputenc}

\newcommand{\fig}[2][]{%
\ifthenelse{\isempty{#1}}
{Fig.~\ref{#2}}
{Fig.~\ref{#2}#1}
}

\begin{document}

\title{Nonlinearity-induced symmetry breaking in a system of two  parametrically driven Kerr-Duffing oscillators}
\author{F. Hellbach}
\address{Fachbereich Physik, Universit{\"a}t Konstanz, D-78457 Konstanz, Germany}
\author{D. De Bernardis}
\address{CNR-INO, care of European Laboratory for Non-Linear Spectroscopy (LENS), Via Nello Carrara 1, Sesto Fiorentino, 50019, Italy}
\author{M. Saur}
\address{Fachbereich Physik, Universit{\"a}t Konstanz, D-78457 Konstanz, Germany}
\author{I. Carusotto}
\address{Pitaevskii BEC Center, CNR-INO and Dipartimento di Fisica, Universit\`a di Trento, I-38123 Trento, Italy}
\author{W. Belzig}
\address{Fachbereich Physik,  Universit{\"a}t  Konstanz, D-78457 Konstanz, Germany}
\author{G. Rastelli}
\address{Pitaevskii BEC Center, CNR-INO and Dipartimento di Fisica, Universit\`a di Trento, I-38123 Trento, Italy}

%
%
\begin{abstract} 
We study the classical dynamics of a system comprising a pair of Kerr-Duffing nonlinear oscillators, which are coupled through a nonlinear interaction and subjected to a parametric drive.
Using the rotating wave approximation (RWA), we analyze the steady-state solutions for the amplitudes of the two oscillators.
For the case of almost identical oscillators, we investigate separately the cases in which only one oscillator is parametrically driven and in which both oscillators are simultaneously driven.
In the latter regime, we demonstrate that even when the parametric drives acting on the two oscillators are identical, the system can transition from a stable symmetric solution to a broken-symmetry solution as the detuning is varied. 
\end{abstract}
	
	

%
%
%
%
\section{Introduction}

Nonlinear dynamics of resonantly driven oscillator modes is a very general theme common to many fields of physics,
from nanomechanical systems to photonics.
In the mechanical context, it has received growing attention in view of metrological applications~\cite{Dykman-book:2012,Cleland-book:2003,Schmid-book:2016,Cleland-Roukes:2002,Lifshitz-Cross:2008,Poot:2012,Rhoads:2010} as well as from a fundamental science perspective. Driven nonlinear nanomechanical systems constitute, in fact, an ideal platform for the investigation of fundamental aspects of nonlinear dynamics and nonequilibrium fluctuations~\cite{ Lifshitz_ron_cross_NanomechanicalResonators, Guttinger:2017is, Chen:2017fl, Bachtold-RMP:2022}. 
In photonics, models involving a few resonantly-driven modes coupled by nonlinear processes provide a powerful framework to describe nonlinear optical processes leading to the mixing of different beams and to the generation of new harmonic and/or sub-harmonic components and are at the heart of devices for quantum and/or nonlinear optics applications~\cite{Lugiato,ButcherCotter,Boyd,WallsMilburn,Drummond} .%

In general, nonlinear coupling between resonant modes is expected to become strong when the ratio between their resonance frequencies 
is an integer (linear resonance 1:n) or rational (parametric resonance 2:n), viz. the so-called internal resonances \cite{Nayfeh:1979}. In particular, an interaction of the kind 2:2, also known as cross-Kerr interaction can be described as the oscillations of the first resonator acting as a parametric frequency modulation on the second resonator.
When internal resonances are not present, the 2:2 nonlinear interaction can still give a dispersive shift of the mode frequencies, which depends on the oscillation amplitude of the driven mode. 
Such a dispersive interaction has been observed in a variety of nanomechanical systems, e.g. doubly clamped beams or strings, clamped nano- or microcantilevers, and in doubly clamped, nanomechanical silicon nitride string resonators
\cite{Westra:2010,Lulla:2012,Matheny:2013,Vinante2014,Mangussi:2016,Cadeddu:2016,Dong:2018,Mathew2018,Gajo:2020}.

In photonics, nonlinear interaction processes are mediated by the nonlinear response of the material medium, in particular second-order $\chi^{(2)}$ and third-order $\chi^{(3)}$ nonlinearities, leading to a variety of phenomena, from optical bistability to parametric amplification and oscillation~\cite{Lugiato,ButcherCotter,Boyd,WallsMilburn,Drummond,Valagiannopoulos9445654,Carusotto:RMP2013}. In the last decades, growing attention has been paid to systems featuring a complex interplay of several nonlinear phenomena at once and/or well-defined discrete modes~\cite{abbarchi2013macroscopic,Wouters:PRB2007,Sarchi:PRB2008}. 
Nowadays, an active frontier is to explore fundamental aspects of non-equilibrium fluctuations around the steady-state of spatially extended systems and shine a light on their relation to fundamental questions of pattern formation in nonlinear systems~\cite{Cross:RMP1993,Fontaine:Nature2022,Claude:arXiv2023,Zamora:PRX2017}.

In this paper, we investigate a simple model based on two Kerr/Duffing oscillators coupled through nonlinear interaction terms, proportional to the product of the squares of the two amplitudes of the two modes $\sim q_1^2 q_2^2$.
Within the rotating wave approximation (RWA), we analyze the possible stationary paired solutions and their dynamical stability in two scenarios: one where only one of the two modes is parametrically driven, and another where both resonators are driven.

When both resonators are parametrically driven
at the same frequency, the paired amplitudes of the two modes show a complex and intriguing behavior which 
is also characterized by the presence of several instability ranges in the frequency detuning.
As a particularly interesting result, we find that when the two resonators have identical parameters and are driven with equal parametric force, the overall symmetry of the system can be spontaneously broken for moderately weak nonlinear interaction. 
This leads to an intriguing multistability phenomenon, featuring steady states with different amplitudes and phases on the two oscillators.

The article is organized as follows.
In Sec.~\ref{sec:2-Model}, we introduce the model and the effective dynamical equations in the rotating frame using the RWA.
In Sec.~\ref{sec:3-one-driven}, we shortly discuss the behavior of the system 
when both resonators have non-zero oscillation amplitude  when only one resonator is driven.
In Sec.~\ref{sec:4-two-driven}, we start analyzing the most intriguing case  when both resonators are simultaneously driven 
and the emergence of symmetry breaking. In the successive Sec.\ref{sec:Symmetric_analytic} and Sec.\ref{sec:Broken_symmetric_numerical}, we respectively give an analytical form for the symmetric solution and, then, a numerical analysis for the broken symmetry case.
Conclusions are finally drawn in Sec.\ref{sec:summary}.

%
%
%
%
\section{The system}
\label{sec:2-Model}
We analyze a system composed by 
two nonlinear resonators of frequency $\omega_n$ with 
$n=1,2$, Kerr-Duffing parameter $\kappa_n$ and damping coefficient $\Gamma_n$, 
on which one applies one or two parametric external drives 
at frequency $2\omega_d$ and of strength $\mu_n$.
We set the two detunings as $\delta\omega_n=\omega_d-\omega_n$.
The two resonators are coupled by a cross-Kerr interaction 
$H_{int}=(\lambda/2) q_1^2 q_2^2$
with $q_n$ the oscillatory amplitudes. 
We refer to the \ref{app:model} for further details.

In the rotating frame and using the RWA, 
the  equations for the scaled complex amplitudes associated with the two resonators oscillating at frequencies $\omega_d$ 
are
%
%
%
%
%
%
%
\begin{equation}
\label{eq:Main-1}
\frac{\partial z_1}{\partial \tau} = i F_1\left(z_1,z_2 \right) \, , \quad 
\frac{\partial z_2}{\partial \tau} = i F_2\left(z_2,z_1 \right)
\end{equation}
%
%
%
%
%
%
with  {$\tau$  the scaled time} and 
the functions $F_n\left(z_n,z_m\right) $  
%
%
%
%
%
%
%
\begin{equation}
F_n\left(z_n,z_m\right)
=
\left( {|z_n|}^2  - \delta\Omega_n   + i \right) z_n
+  \delta \Omega_{T,n} z_n^* 
 +  g  \left(2  {|z_m|}^2 z_n + z_m^2 z_n^{*} \right) \, .
\label{eq:Main-2}
\end{equation}
for $n=1$ and $m=2$ and viceversa.

Here $\delta\Omega_n $ corresponds to the scaled detuning for the resonator $n$ 
and 
$\delta\Omega_{T,n} $ is proportional to the strength of the parametric drive.
It also corresponds to the (scaled) frequency threshold for the self-sustained oscillatory motion of the resonator $n$ 
in the limit of vanishing damping and without interaction. 
The parameter $g$ represents the scaled coupling strength of the nonlinear interaction.

The stationary solution are obtained by requiring  $\partial z_1/\partial \tau = \partial z_2/\partial \tau = 0$,
namely by solving the two coupled nonlinear equations
%
%
%
%
%
%
%
\begin{equation}
F_1\left( \bar{z}_1,\bar{z}_2\right) = F_2\left(\bar{z}_2,\bar{z}_1\right) =0 \, .
\label{eq:Fstat}
\end{equation}
 {
The numerical solutions for the nonlinear coupled Eqs.~(\ref{eq:Fstat}) 
were obtained using a Mathematica code based on the NSolve function with optimized parameters.
}

We analyze the dynamical stability of these solutions by analyzing the small fluctuation around these points 
 $z_n \simeq \bar{z}_n  +\delta z_n$, namely, we consider the equations for the vector of the fluctuations
%
%
%
%
%
%
%
\begin{equation}
\frac{\partial }{\partial \tau}
\left( 
\begin{array}{c}
\delta z_1 \\
\delta z_1^* \\
 \delta z_2 \\
 \delta z_2^*
\end{array}
\right)
=
i 
\left(
\begin{array}{cccc}
a_1		&	b_1		&	c	& 	d		\\
-b^*_1		&	-a_1		&	-d^*	& 	-c		\\
c		&	d		&	a_2	& 	b_2		\\
-d^*		&	-c		&	-b_2^*	& 	-a_2		
\end{array}
\right)
\left(
\begin{array}{c}
\delta z_1 \\
\delta z_1^* \\
 \delta z_2 \\
 \delta z_2^*
\end{array}
\right)
\label{eq:du}
\end{equation}
with the coefficients evaluated at the stationary solutions 
%
%
%
%
%
\begin{eqnarray}
a_1 & = 2 {|\bar{z}_1|}^2 - \delta\Omega_1  +2 g   {|\bar{z}_2|}^2 + i 	\, , \\
a_2 & = 2 {|\bar{z}_2|}^2 - \delta\Omega_2  +2 g   {|\bar{z}_1|}^2	 + i	\, ,	\\
b_1 & = \bar{z}_1^2 + \delta\Omega_{T,1}  + g   \bar{z}_2^2 			\, ,\\
b_2 & = \bar{z}_2^2 + \delta\Omega_{T,2}  + g   \bar{z}_1^2			\, ,\\
c	&= 2 g \left( \bar{z}_1^* \bar{z}_2 +  \bar{z}_1 \bar{z}_2^* \right)	\, , \\
d	&=2 g  \bar{z}_1 \bar{z}_2 \, .
\end{eqnarray}

When the real parts 
of all 
eigenvalues of the matrix appearing in Eq.~(\ref{eq:du}) 
 - including the factor $i$ - 
are negative, the stationary solution is dynamically stable against noise.
In the limit of small damping, which corresponds to neglect the factor $i$ in the coefficients $a_1$ and $a_2$, the solutions are dynamically stable simply when the real parts of the eigenvalues vanish.

Hereafter we  focus our analysis on the strong nonlinear regime in which the motion is dominated by the nonlinear interaction such that we can neglect 
the damping force, which is valid in the limit  $3 \kappa A_n^2/(8\omega_d) \gg \Gamma$ with $A_n$ the amplitude of the oscillatory motion 
of the resonator $q_n(t) = A_n \cos(\omega_d t + \theta_n)$, $\kappa$ is the Kerr-Duffing parameter, $\Gamma$ is the damping and 
$\omega_d$ is half of the drive frequency (see    \ref{app:model}). 
This approximation is not valid when we are close to the parametric threshold when the resonator starts to oscillate at frequency $\omega_d$ 
and the amplitude can be small enough to violate the previous condition.
However, the effect of the damping is the renormalization of the threshold. 
Even in the presence of a nonlinear interaction, we have numerically benchmarked that small damping simply results in a minor renormalization or shift of the critical thresholds when new solutions emerge or existing solutions vanish with varying detuning.

Before analyzing the solutions in which both resonators oscillate, $\bar{z}_1\neq0$ and $\bar{z}_2\neq0$, we discuss shortly 
the simple cases in which one of the two resonators is at rest.
We refer to the   \ref{app:g=0_z=0} and  \ref{app:z2=0} for a more extended analysis. 

First of all, for a single parametrically driven nonlinear resonator (uncoupled case $g=0$), one has the following (stable) steady-state solution 
for $\bar{z}$: the resonator starts to oscillate at frequency $\omega_d$ above the detuning threshold $ \delta\Omega> - \delta\Omega_{T,1}$ 
and it is entirely out of phase with respect to the drive, in the limit of vanishing damping (see   \ref{app:g=0_z=0}).
The trivial solution $\bar{z}^{(0)}_{1}=0$ is unstable in the detuning range $|\delta \Omega | < \delta\Omega_{T,1}$.

When we switch on the interaction, we discuss the stationary pair solutions.
In particular, the trivial solution $\left( \bar{z}_{1} =0 , \bar{z}_{2}=0 \right) $ is still a stationary solution and  
unstable for $| \delta\Omega_i |< \delta\Omega_{T,i}$ as for the case $g=0$.
On the other hand, for the pair solutions with the trivial solution for one of the two resonators $\bar{z}_n=0$, the equations for $\delta z_1$ and $\delta z_2$ become decoupled (see Eq.~(\ref{eq:du}), the coefficients $c$ and $d$ vanish) 
but 
they are still correlated as 
the finite value of the amplitude of the driven mode $\bar{z}_n$ affects the fluctuation of the second non-driven mode  $\delta z_m$ 
and 
therefore the range of stability can be different from the noninteracting case,  see   \ref{app:z2=0}.

\section{Finite amplitude solutions for one driven resonator}
\label{sec:3-one-driven}
We now analyze the system when the external drive is applied only to the first resonator  $(\delta\Omega_{T,2}= 0)$.
In   \ref{app:single-drive}, we give a detailed analysis of this regime, in which we report the analytical solutions in table \ref{Tab:1}.
The different possible solutions are associated with different ranges of the coupling strength $g$, in particular 
in the ranges $0<g<1/3$, $1/3<g<1$ and $g>1$.
We checked that 
the numerical results are in good agreement with the analytical solutions 
shown in   \ref{app:single-drive}.

%
%
%
\begin{figure}[h]
\includegraphics[width=0.5\linewidth]{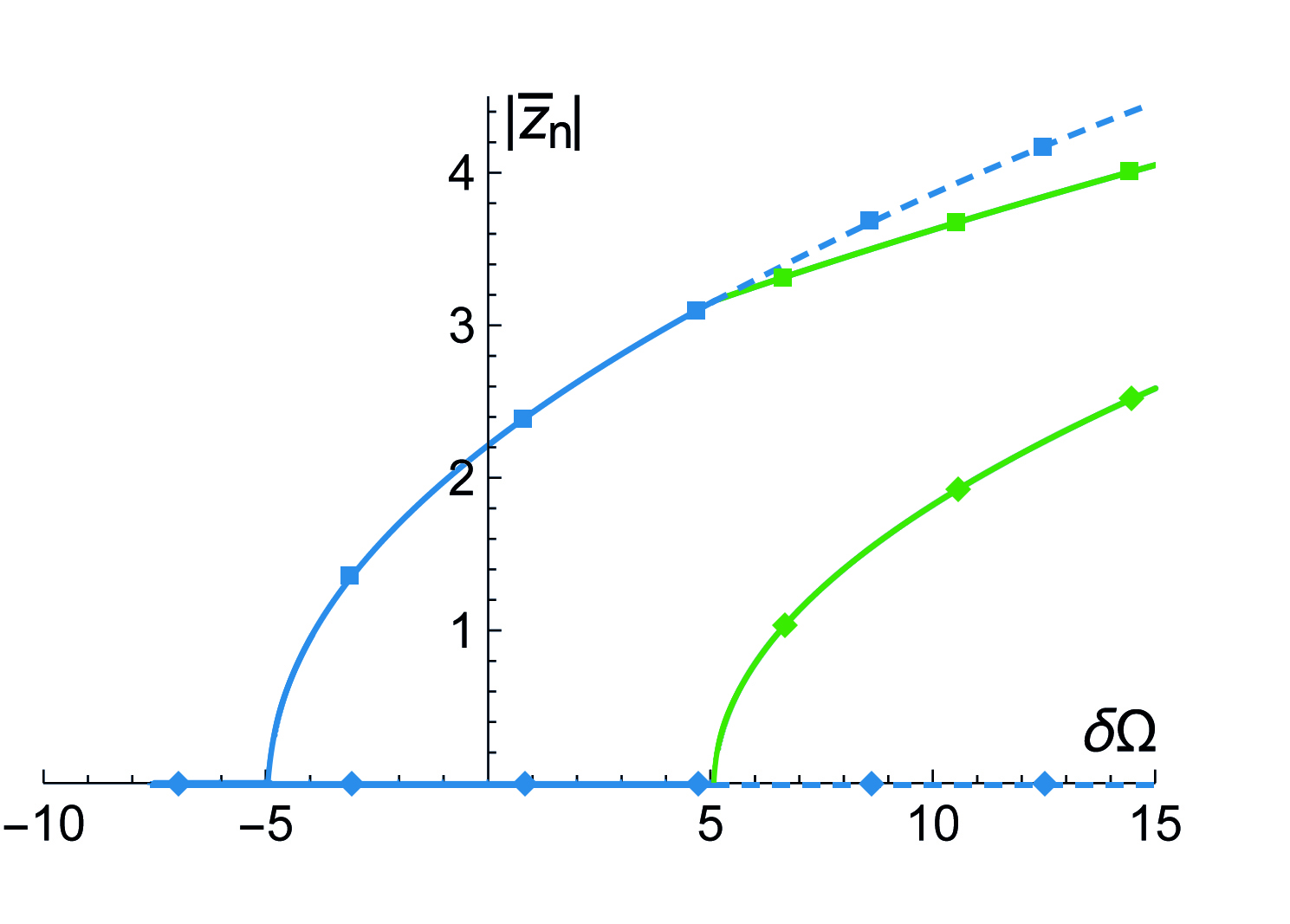}
\caption{
Example of the behavior for a single-driven mode. 
Legend: The amplitudes of the two different resonators are distinguished by the symbol, i.e., $\blacksquare$ for $|\bar{z}_1|$, 
$\blacklozenge$
for $|\bar{z}_2|$. Lines with the same color show solution pairs of the two resonators. Solid lines correspond to stable solutions. Dashed lines correspond to unstable solutions.
Parameters:  $g=1/2$, $\delta\Omega_{T,1}=5$ and $\delta\Omega_{T,2}=0$. The blue lines show the solution pairs with only one oscillating resonator while the other one is at rest. }
\label{fig:1}
\end{figure}

In short, when only the first mode is driven, the cross-Kerr interaction changes the system's stability range even when we have the trivial solution for the second non-driven mode. 
Furthermore, the first resonator can drive the state of the second resonator in an oscillatory motion.

We give an example of results for this regime in Fig.~\ref{fig:1} (numerical results) in which we plot stable pair solutions (solid lines) and unstable ones (dashed lines).
In particular, we notice that, for the stable solutions,  the second non-driven resonator starts to oscillate for a given threshold detuning.
This leads to a discontinuity in the amplitude curve of the driven resonator. 
Changing the parameters of the system, we also find that 
the curve for the amplitude of the second resonator can have 
a discontinuity in which the amplitude $\bar{z}_2$ has a jump from zero to a finite value.

\section{Finite amplitude solutions  for two driven resonators}
\label{sec:4-two-driven}
In this section, we analyze the case of both nonlinear resonators being parametrically driven.
We consider the degenerate case, with $\omega_1=\omega_2$ such that the detuning frequencies are the same 
$\delta\Omega_1=\delta\Omega_2=\delta\Omega$.\\

\subsection{Numerical analysis for asymmetric drive $\delta\Omega_{T,1}\neq \delta\Omega_{T,2}$}
We start the analysis when the strengths of the two parametric drives are different $\delta\Omega_{T,1 } \neq \delta\Omega_{T,2}$. 
For this case, we  discuss some simple limits, in a qualitative manner, that we have tested numerically and show 
an example of results.

First of all, 
in the limit of weak coupling $g\ll1$,  the solutions are connected qualitatively to uncoupled solutions for two noninteracting modes, as expected.

\begin{figure}[b]
\includegraphics[width=0.5\linewidth]{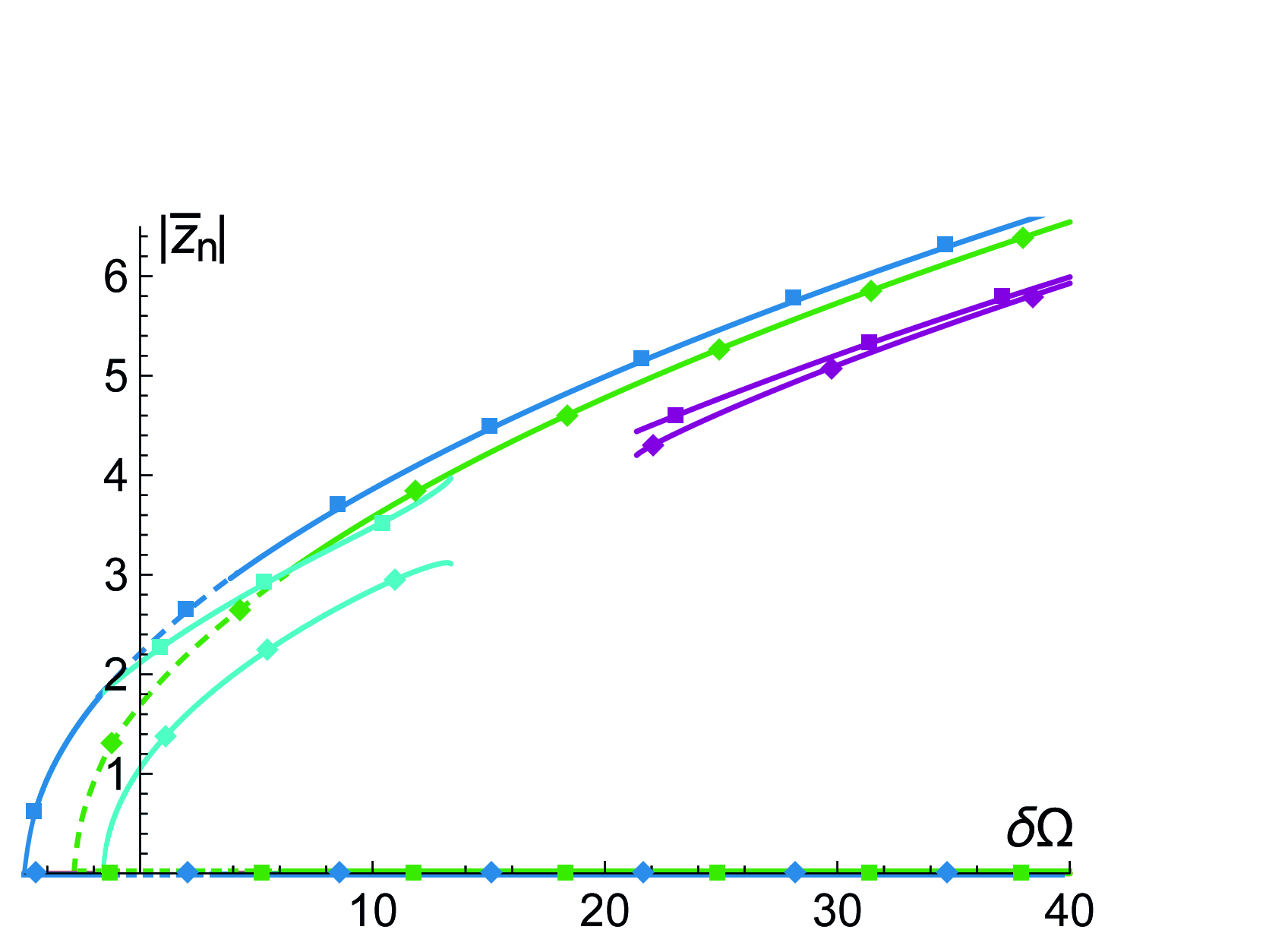} 
\includegraphics[width=0.5\linewidth]{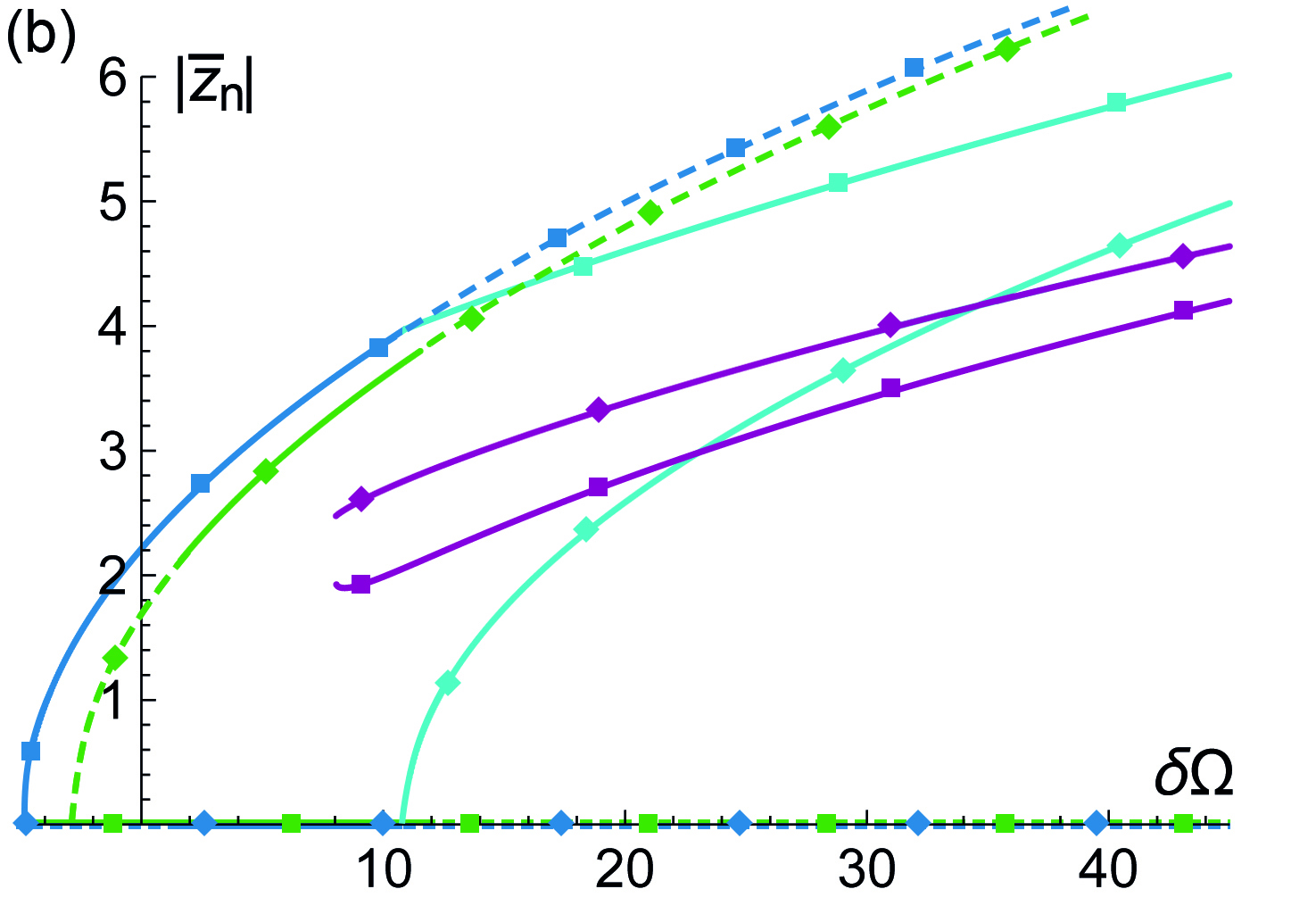}
\caption{
Example of the behavior of the system for two driven modes with different drive strengths. 
Legend: The amplitudes of the two different resonators are distinguished by the symbol, i.e. $\blacksquare$ for $|\bar{z}_1|$,
$\blacklozenge$ 
for $|\bar{z}_2|$.  Lines with the same color show solution pairs of the two resonators. Solid lines correspond to stable solutions. Dashed lines correspond to unstable solutions.
The green and blue lines are the paired solutions with $\bar{z}_1=0$ or $\bar{z}_2=0$.
Parameters: 
{\bf (a)} $\delta\Omega_{T,1}=3$, $\delta\Omega_{T,2}=5$, $g=1/8$.
{\bf (b)} $\delta\Omega_{T,1}=10/3$, $\delta\Omega_{T,2}=5$, $g=1/2$.
}
\label{fig:2}
\end{figure}

On the other hand, for $\delta \Omega_{T,1} \gg \delta \Omega_{T,2} $, for moderate force acting on the first resonator $\delta\Omega_{T,1}\gg 1$ and 
not so small coupling $g$,  the solutions are connected qualitatively to ones  
obtained for the case of a single-driven mode, see 
Sec. \ref{sec:3-one-driven}.
One can estimate this regime in the small damping limit in which we have 
${|z_{1}|}^2 \approx \delta \Omega_{T,1}$  at small detuning.
Hence the effects of the first resonator dominate on the second drive for $g   \delta\Omega_{T,1} \gg  \delta\Omega_{T,2} $.
In such regime  
the interaction between the two resonators dominates over the second drive, the amplitude of the second - weakly driven - resonator is  
determined by its parametric interaction with the first resonator 
in a way qualitatively similar to the results discussed in 
Sec. \ref{sec:3-one-driven}.

The more interesting regime occurs when the strength of the two drives are comparable, and the interaction strength is also not so small.
An example of results in such a regime  is shown 
in Fig.~\ref{fig:2} for  $g=1/8$ and  $g=1/2$.
Here the blue and green lines are the paired solutions with $\bar{z}_2=0$ or $\bar{z}_1=0$, respectively.  
The solid lines show the stable solutions.
As there are many unstable solutions, 
we include only a few examples of them 
in the figure corresponding to the dashed lines.

For $g=1/8$ , we have stable solutions of finite amplitudes,  $\bar{z}_1 \neq 0$ and
$\bar{z}_2 \neq 0$, 
(turquoise lines) up to some detuning, after that 
they disappear, and two new stable solutions emerge again at larger detuning (purple lines).
The system has multistability when we regard the phase since the modulus of the solutions shown in Fig.~\ref{fig:2} corresponds to solutions of different phases.
One can show that, in total, there are four different paired solutions corresponding to the four possible combinations of the phases.

For the case $g=1/2$, the system shows a multistability even for the modulus, 
as shown Fig.~\ref{fig:2}b, in which two paired solutions of different amplitudes (turquoise lines and purple lines) are stable in the same detuning range.
In this case, the system admits 
eight stable  paired solutions, e.g. four solutions for 
a given pair of amplitudes.

\subsection{Numerical analysis for symmetric drive  $\delta\Omega_{T,1} = \delta\Omega_{T,2} $}

\begin{figure}[b]
\includegraphics[scale=0.4]{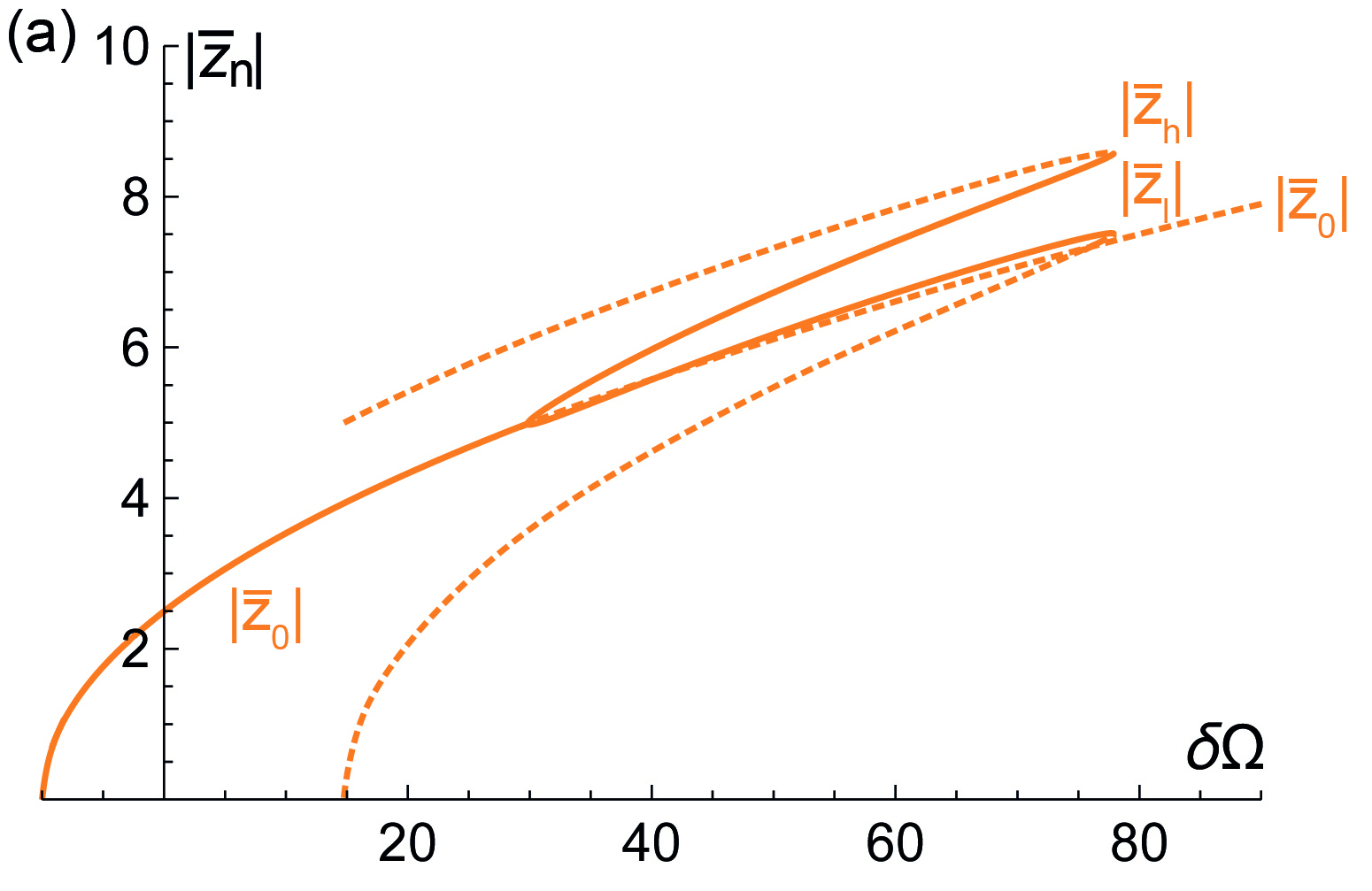} \includegraphics[scale=0.4]{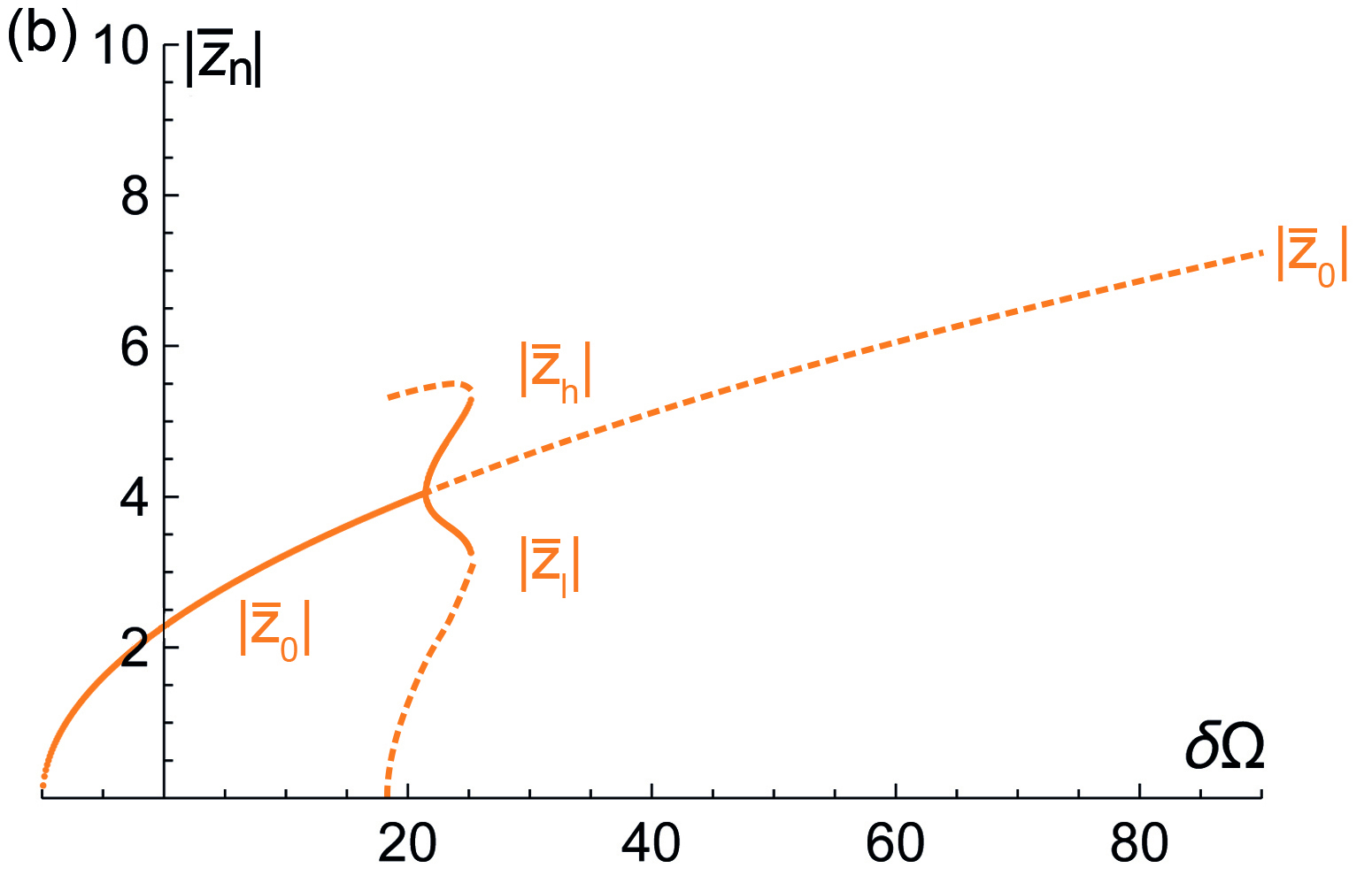}\\
\includegraphics[scale=0.4]{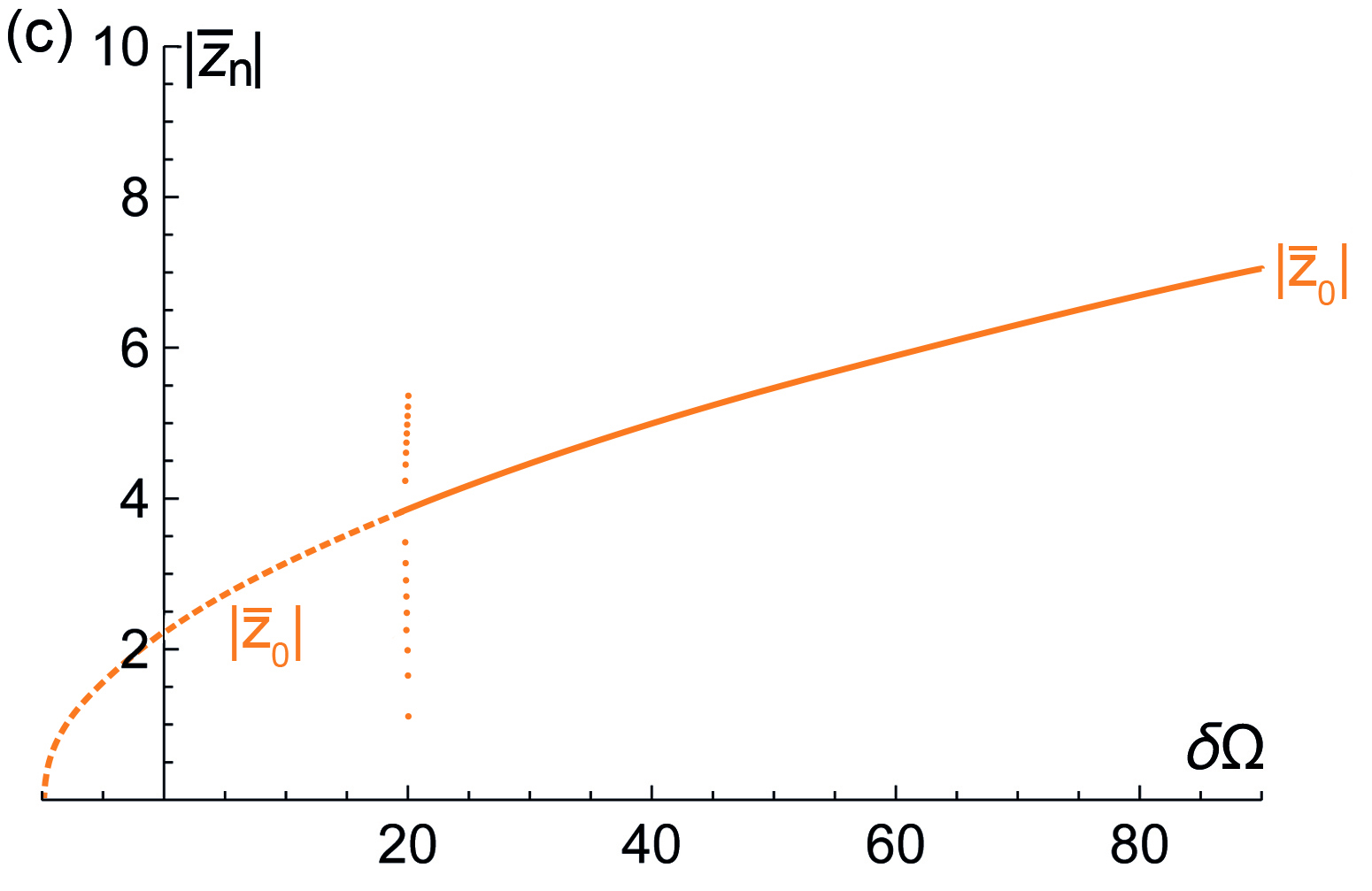}\includegraphics[scale=0.4]{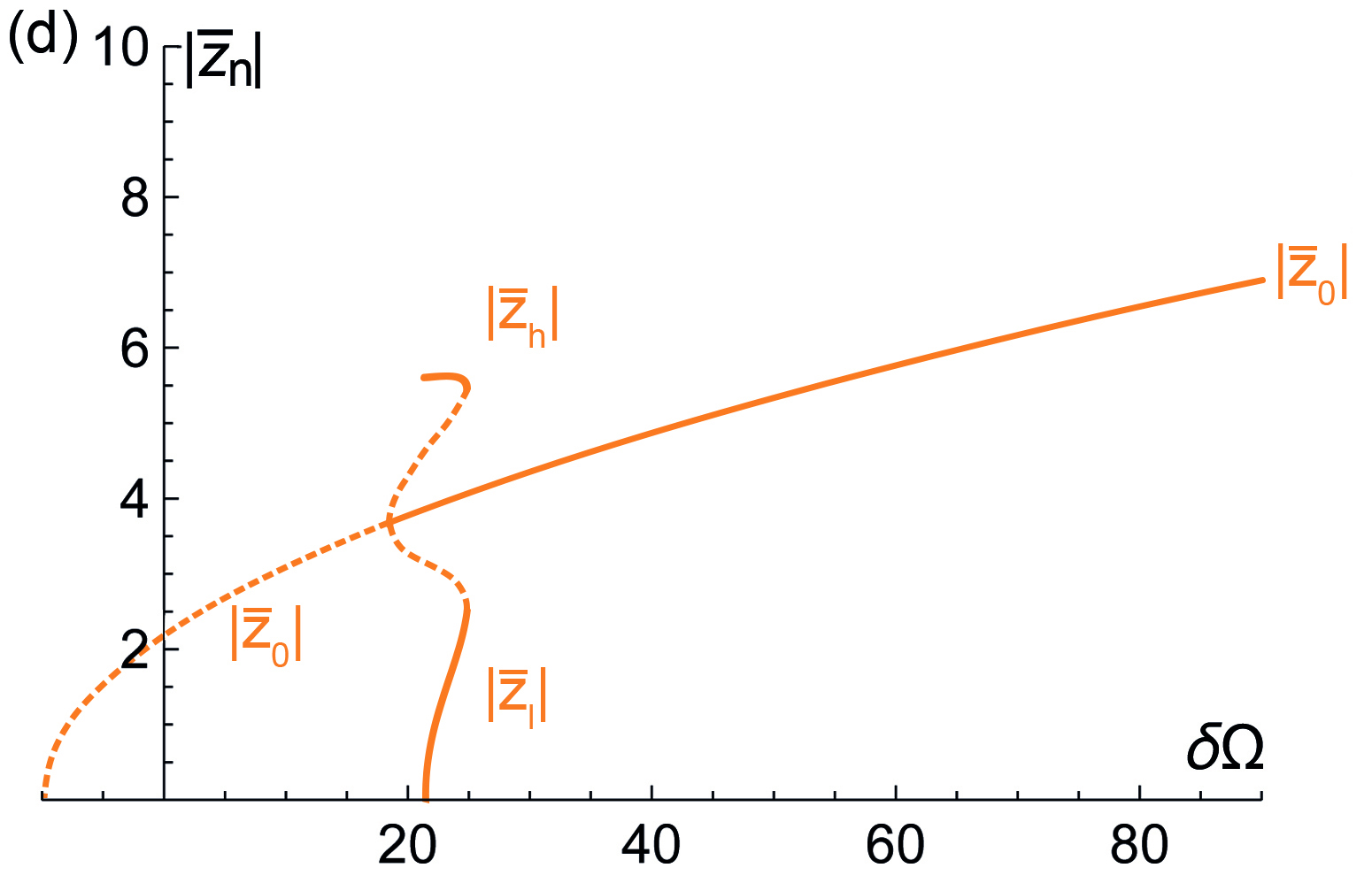}
\includegraphics[scale=0.4]{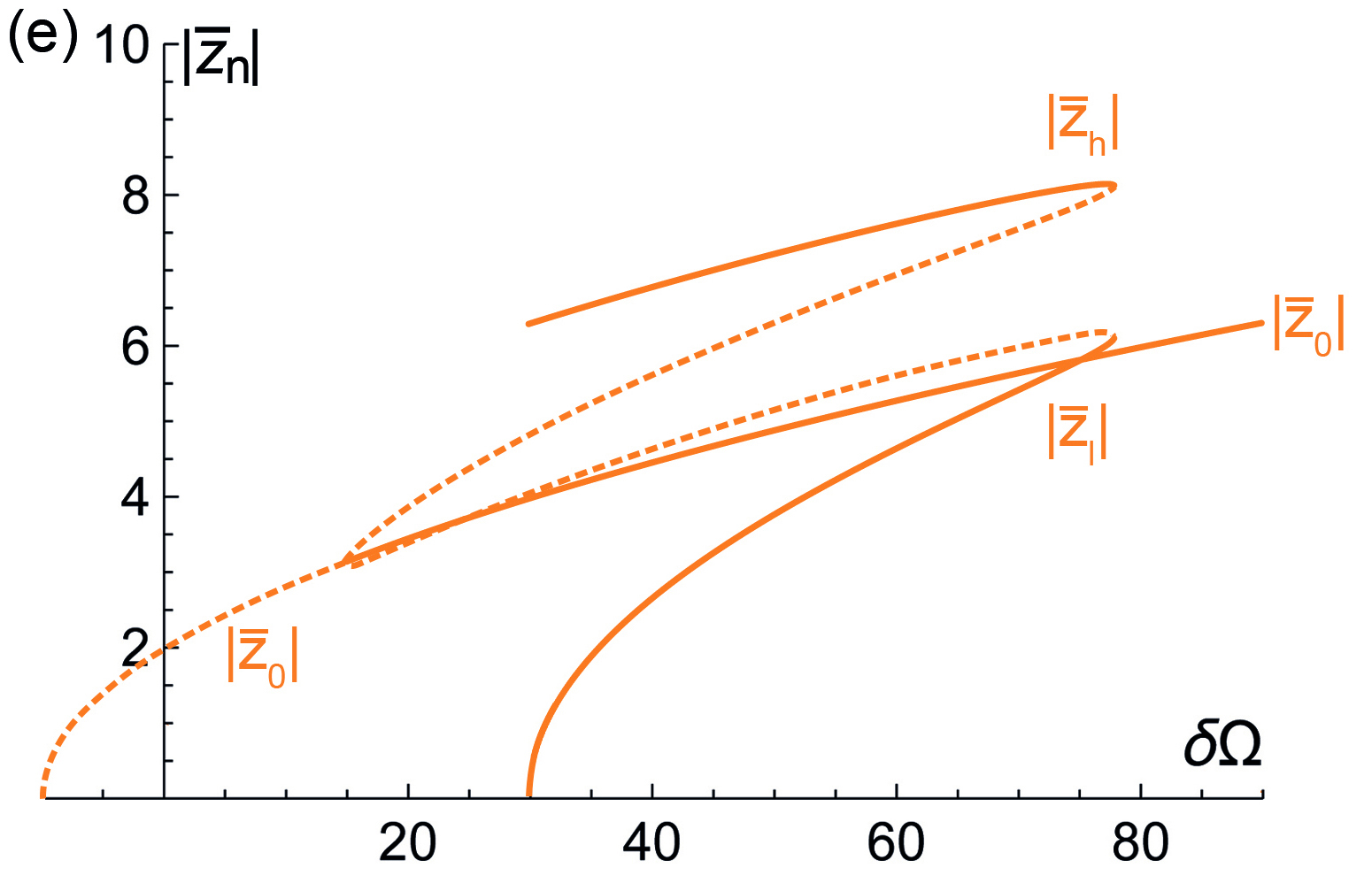}
\caption{
Example of the behavior of the system for two driven modes with symmetric drive.
The  line $|\bar{z}_0|$ is the symmetric solution, namely the solution with the same amplitude for both resonators.
The solid lines are the stable solutions, whereas the dashed lines correspond to the unstable ones.
Parameters: $\delta\Omega_{T,1}=\delta\Omega_{T,2}=10$; 
{\bf (a)}  $g=1/5$; 
{\bf (b)} $g=3/10$; 
{\bf (c)} $g=101/300$;
{\bf (d)} $g=11/30$;
{\bf (e)} $g=1/2$.
}
\label{fig:3}
\end{figure}

As the most interesting case, we analyze the case in which the strength of the two drives is the same $\delta\Omega_{T,1}=\delta\Omega_{T,2}$. 
In this case, the system is fully symmetric.
An example of results in such a regime  is shown in Fig.~\ref{fig:3} 
for $g=1/5$, $g=3/10$ and $g=1/2$, in the limit of negligible damping.

The symmetry of the equations leads to a natural symmetric solution in which the two resonators have the same oscillatory amplitude $|\bar{z}_1|=|\bar{z}_2 | = |\bar{z}_0|$.
We find that this solution is stable for different ranges, 
as indicated in Fig.~\ref{fig:3}.
Although the modulus is the same, 
the system has a multistability related to the combinations of 
all possible phases of the solutions (similar to the results of the previous section).
Denoting the two possible states for each resonator as 
$|\bar{z}_0| e^{i\varphi_{0,+}}$ and 
$|\bar{z}_0| e^{i\varphi_{0,-}}$, 
with $|\varphi_{0,+}-\varphi_{0,-}|=\pi$, 
we obtain  in total four possible solution pairings 
depending on whether the two resonators have either different or equal phases. 
They correspond to  ($\varphi_{0,+}$,$\varphi_{0,+}$), 
($\varphi_{0,+}$,$\varphi_{0,-}$), 
($\varphi_{0,-}$,$\varphi_{0,+}$), 
($\varphi_{0,-}$,$\varphi_{0,-}$).

 {
When the symmetric solutions become unstable in Fig.~\ref{fig:3}a and
Fig.~\ref{fig:3}b, we observe solutions with broken symmetry, where the two resonators exhibit different oscillatory amplitudes. We refer to these solutions as $|\bar{z}_{l}|$ (low amplitude) and $|\bar{z}_{h}|$ (high amplitude). In this regime, the system exhibits multistability with respect to the pair of solutions: if one resonator is in the low amplitude state $|\bar{z}_{l}|$, the other must be in the high amplitude state $|\bar{z}_{h}|$.
}

When we increase the coupling, we observe that at $g=1/3$ 
the region of stability of the symmetric solution changes 
discontinuously.
For $g<1/3$ the symmetric solution is stable 
from $\delta\Omega > - \delta\Omega_T$ up to a critical detuning $\delta\Omega < \delta\Omega_{c,S}$, see Fig.~\ref{fig:3}a-b, 
whereas it becomes 
stable 
at large detuning $\delta\Omega > \delta\Omega_{c,S}$ 
for $g>1/3$, 
see Fig.~\ref{fig:3}c-d-e . 

 {
In Fig.~\ref{fig:3}c-d-e, for $g>1/3$, the (stable) symmetric solution coexists with the (stable) broken symmetry solution at large $\delta\Omega > \delta\Omega_{c,S}$.
}

We have numerically calculated the critical value $g$ at which the system with 
broken symmetry switches from 
the solution types of Fig.~\ref{fig:3}a-b
to the  solution types of Fig.~\ref{fig:3}c-d-e, 
at different
values of the drive strength $\delta\Omega_{T}$.
Within the range of parameters explored in our numerical 
solutions, we find that the critical coupling at which 
the discontinuity occurs is $g_c \approx 1/3$ and it is 
independent of the drive strength $\delta\Omega_T$.
The latter only affects the critical detunings  
$\delta\Omega_{c,S}$ at different $g$.
In Fig. \ref{fig:3}c in particular we see an example extremely close to the critical coupling.

For $g<1/3$, at $\delta \Omega = \delta \Omega_{c,S}$, 
we have a pitchfork bifurcation of the stable solutions: 
for $\delta \Omega > \delta \Omega_{c,S}$, 
the symmetric solution is unstable, and we have a solution with broken symmetry in amplitudes 
$|\bar{z}_1| \neq |\bar{z}_2|$,
see Fig.~\ref{fig:3}a and Fig.~\ref{fig:3}b.
For $g>1/3$, the symmetric solution becomes stable for 
$\delta \Omega > \delta \Omega_{c,S}$ 
whereas the broken-symmetry solution results are stable at increasingly larger detuning, see Fig.~\ref{fig:3}d-e, with one of them appearing at finite amplitude in a discontinuous way.

 {
At the critical point $g=1/3$ the broken symmetry solutions become all unstable, organized along a straight line, as visible in Fig. \ref{fig:3}c for $g=101/300$. However the presence of this large number of unstable solution in a small range of detuning makes the numerical algorithm very slow while approaching $g\rightarrow 1/3$ making very difficult to compute the case exactly at the critical coupling. 
}

\section{Symmetric solutions for symmetric drive}
\label{sec:Symmetric_analytic}

In the limit of vanishing damping, the symmetric solutions take 
the form 
 {
\begin{equation}
z_0 = \pm i \, \sqrt{ \frac{\delta\Omega+\delta\Omega_T}{1+3g} }
\end{equation}
}
where we set $\delta\Omega_T=\delta\Omega_{T,1}=\delta\Omega_{T,2}$.
Then the coefficients of the matrix associated to the fluctuations 
Eq.~(\ref{eq:du}) around the stationary points are real and read
\begin{eqnarray}
a & = 2 \left(1 + g \right) {\left| z_0 \right|}^2- \delta\Omega \, , \\
b & = - \left(1 + g \right) {\left| z_0 \right|}^2 + \delta\Omega_T	\, ,\\
c &= \pm 4 g {\left| z_0 \right|}^2	\, , \\
d &= \mp 2 g {\left| z_0 \right|}^2	\, ,
\end{eqnarray}
with the sign $\pm$  $(\mp)$  depending on the phases of the paired solution, e.g.  the same $\bar{z}_1=\bar{z}_2$ or different phase $\bar{z}_1=-\bar{z}_2$.
Setting  
$\delta z_n= \delta Q_n - i \delta P_n$ and 
changing the variables 
$\delta Q_{\Sigma} =\delta Q_1+ \delta Q_2$,
$\delta Q_{\Delta} =\delta Q_1-  \delta Q_2$,
$\delta P_{\Sigma} =\delta P_1+  \delta P_2$,
$\delta P_{\Delta} =\delta P_1-  \delta P_2$, 
one can obtain the expansion of the Hamiltonian around the symmetric stable solution as the same of two quadratic Hamiltonian
\begin{equation}
H_\mathrm{RWA}^{(2)} 
= 
H_{\Sigma}
+  
H_{\Delta}
\, ,
\end{equation}
with 
\begin{eqnarray}
H_{\Sigma}
&=
\frac{1}{2} \Omega_{P,\Sigma}
\,\, \delta P_{\Sigma}^2
+
\frac{1}{2} \Omega_{Q,\Sigma}
\,\, \delta Q_{\Sigma}^2 \, , \\
H_{\Delta}
& =
\frac{1}{2} \Omega_{P,\Delta}
\,\, \delta P_{\Delta}^2
+
\frac{1}{2} \Omega_{Q,\Delta}
\,\, \delta Q_{\Delta}^2 \, ,
\end{eqnarray}
in which the conjugate variables are 
$\left( \delta Q_{\Sigma}, \delta P_{\Sigma} \right)$
and
$\left( \delta Q_{\Delta}, \delta P_{\Delta}  \right)$ with the following dynamical equations 
\begin{eqnarray}
&
\delta \dot{P}_{\Sigma}  = \partial H_{\Sigma}/ \partial \delta Q_{\Sigma} 
\, , \quad 
\delta \dot{Q}_{\Sigma} = -\partial H_{\Sigma}/ \partial 
\delta P_{\Sigma} \, ,\\
&
\delta \dot{P}_{\Delta} = \partial H_{\Delta}/ \partial 
\delta \dot{Q}_{\Delta} 
\, ,  \quad 
\delta \dot{Q}_{\Delta} 
= -\partial H_{\Delta}/ 
\delta \dot{P}_{\Delta} \, .
\end{eqnarray}
The system is dynamically stable as long as the product of the two 
coefficients in each quadratic Hamiltonian is positive, namely
$\Omega_{P,\Sigma} \Omega_{Q,\Sigma} >0$ 
and 
$\Omega_{P,\Delta} \Omega_{Q,\Delta}  >0$.\\

For example, for $\bar{z}_1=\bar{z}_2$, the coefficients are
\begin{eqnarray}
\Omega_{P,\Sigma} &= 2 \left( \delta\Omega+\delta\Omega_T \right)  \, ,\\
\Omega_{Q,\Sigma} &= 2 \delta\Omega_T \, , 
\end{eqnarray}
and
\begin{eqnarray}
\Omega_{P,\Delta} &= 2\, 
\frac{1-3g}{1+3g}
\left( \delta\Omega+\delta\Omega_T \right)  \, , \label{eq:final_1}
\\
\Omega_{Q,\Delta}
&=
2 \, \frac{\delta\Omega_T}{1+3g}
\left[
1- g \left( 2\frac{\delta\Omega}{\delta\Omega_T} - 1\right)
\right] \, .
\label{eq:final_2}
\end{eqnarray}

The quadratic Hamiltonian around the stationary point  $\bar{z}_1=-\bar{z}_2$ has a similar form which can be written 
by exchanging the coefficient  
$\Omega_{P,\Sigma}$ 
with 
$ \Omega_{P,\Delta} $
and 
the coefficient 
$ \Omega_{Q,\Sigma} $ 
with 
$ \Omega_{Q,\Delta} $.

For positive detuning $\delta\Omega>0$, 
we observe that only the coefficient  Eq.~(\ref{eq:final_1}) 
and Eq.~(\ref{eq:final_2}) can change the sign by varying the detuning 
and the coupling constant.
In particular  Eq.~(\ref{eq:final_1}) is positive 
for $g<1/3$. 
The frequency Eq.~(\ref{eq:final_2}) is always positive in the range 
$0<g<1$ for $\delta \Omega<\delta\Omega_T/2$ whereas 
for $\delta \Omega > \delta\Omega_T/2$ we have the condition 
\begin{equation}
g < \frac{1}{2 \frac{\delta\Omega}{\delta\Omega_T} - 1} 
\,  \qquad \mbox{for} \quad \delta \Omega > \frac{\delta\Omega_T}{2} 
\label{eq:final3} \,. 
\end{equation}
In conclusion, when both Eq.~(\ref{eq:final_1}) 
and Eq.~(\ref{eq:final_2}) are positive or negative, the system is stable: 
  the critical line $g_c=1/3$ and Eq.~(\ref{eq:final3}) determines the stability phase diagram shown in Fig.~\ref{fig:5} for the symmetric solution.

\section{Broken-symmetry solutions for symmetric drive}
\label{sec:Broken_symmetric_numerical}
We now discuss the behavior of the broken symmetry solutions.
At the pitchfork bifurcation, as shown in Fig.~\ref{fig:3}(a) and 
Fig.~\ref{fig:3}(b), we have two amplitudes  which we denote $|\bar{z}_{h}|$ and $|\bar{z}_{l}|$.
 {As explained previously, }
the system has multistability characterized by the possibility of exchanging the amplitudes of the two resonators, i.e. 
we have a state with 
$|\bar{z}_{1}|=|\bar{z}_{h}|$ and $|\bar{z}_{2}| = |\bar{z}_{l}|$  and a different state with 
$|\bar{z}_{1}|=|\bar{z}_{l}|$ and $|\bar{z}_{2}| = |\bar{z}_{h}|$.\\

Furthermore, even if we fix the modulus of the amplitudes of the two resonators, for example, $|\bar{z}_{1}|=|\bar{z}_{h}|$ and $|\bar{z}_{2}| = |\bar{z}_{l}|$,
this state is still characterized by multistability owing to the different phases 
associated to each solution $\bar{z}_{h}$ and $\bar{z}_{l}$.
More explicitly, 
we have four possible states given by  four possible combinations of the phases 
\begin{eqnarray}
(1) \qquad \bar{z}_1= |\bar{z}_{h}| e^{ i \varphi_{h,+}} \, \,\,\, 
\bar{z}_2= |\bar{z}_{l}| e^{ i \varphi_{l,+}} \, , \\
(2) \qquad \bar{z}_1= |\bar{z}_{h}| e^{ i \varphi_{h,+}} \, \,\,\, 
\bar{z}_2= |\bar{z}_{l}| e^{ i \varphi_{l,-}} \, , \\
(3) \qquad \bar{z}_1= |\bar{z}_{h}| e^{ i \varphi_{h,-}} \, \,\,\, 
\bar{z}_2= |\bar{z}_{l}| e^{ i \varphi_{l,+}} \, , \\
(4) \qquad \bar{z}_1= |\bar{z}_{h}| e^{ i \varphi_{h,-}} \, \,\,\, 
\bar{z}_2= |\bar{z}_{l}| e^{ i \varphi_{l,-}} \, .
\end{eqnarray}
with $|\varphi_{h,+}-\varphi_{h,-} |=| \varphi_{l,+}-\varphi_{l,-} |= \pi$.

\begin{figure}[t]
\begin{flushleft}
\includegraphics[width=0.75\linewidth]{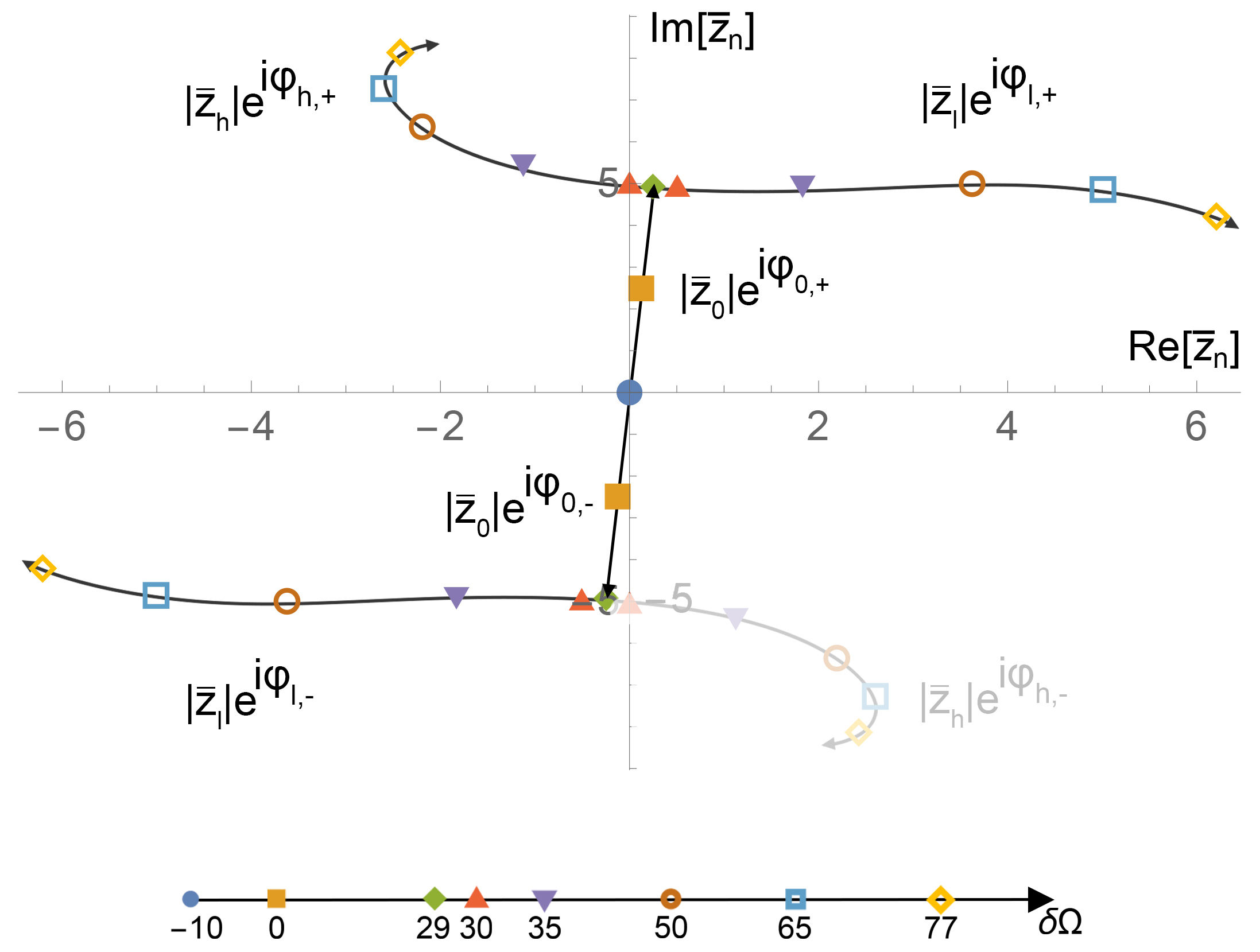}
\end{flushleft}
\caption{
An example of the behavior of the system in the presence of a broken-symmetry solution corresponding to the parameters shown in Fig.~\ref{fig:3} (a).
Parameters:  $g=1/5$, $\delta\Omega_{T,1}=\delta\Omega_{T,2}=10$. 
The stable solutions for the two amplitudes at different values of the detuning are shown. Before the bifurcation, we have the symmetric solutions $|\bar{z}_0|$, which are placed on a diagonal through the origin: the two modes can have either 
the same phase or a phase shift by $\pi$ (see text). 
After the bifurcation, we obtained different stable paired solutions.
For example 
the solution $z_1 = |\bar{z}_{h}| e^{i \varphi_{h,+}}$  
in quadrant II can be paired either with 
 $z_2 = |\bar{z}_{l}| e^{i \varphi_{l,+}}$  in quadrant I 
 or 
 with 
 $z_2 = |\bar{z}_{l}| e^{i \varphi_{l,-}}$  in quadrant III, 
  { but not with the solution 
 $z_2 = |\bar{z}_{h}| e^{i \varphi_{h,-}}$ 
 (shaded region)}. 
 }
\label{fig:4}
\end{figure}

We summarize the behavior of the paired solutions 
in 
Fig.~\ref{fig:4} in which we plot 
the complex amplitudes of Fig.~\ref{fig:3}a 
parametrically as a function of the detuning for the case $g=1/5$.

%
%
%
\begin{figure}[t]
\includegraphics[scale=1.0]{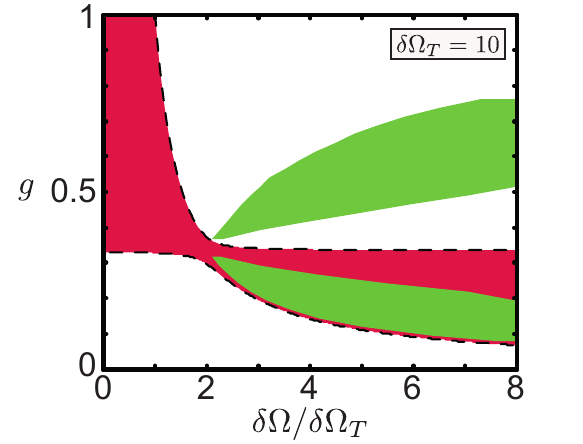}
\caption{Phase diagram for the symmetric resonators with equal resonant frequency and 
drive strength $\delta\Omega_T=10$. The symmetric solutions $|\bar{z}_1|=|\bar{z}_2|$  are stable (unstable) in the white (red) region. The broken symmetry solutions are stable in the green region. }
\label{fig:5}
\end{figure}

In Fig.~\ref{fig:5}, we show the complete phase diagram. The white areas represent regions where symmetric solutions are dynamically stable. In contrast, red areas indicate regions of dynamic instability for these solutions. Green areas denote the existence and dynamic stability of broken-symmetry solutions. When green overlaps with red, the broken-symmetry solution is the only stable configuration. Conversely, where green overlaps with white, both symmetric and broken-symmetry solutions remain dynamically stable.

We found 
the stability regions for the broken-symmetry solutions numerically, whereas the symmetric solutions are obtained from analytical 
formulas discussed in the previous section. 
A similar phase diagram occurs if we vary the symmetric drive strength $\delta\Omega_{T,1}=\delta\Omega_{T,2}$ whereas the critical coupling at $g_c \approx 1/3$ 
remains unchanged.

Given the parametric form of the driving and of the coupling, one must not forget that solutions where at least one oscillator is at rest, $z_1=0$ or $z_2=0$, are always possible.

\section{Summary}
\label{sec:summary}
In conclusion, we have studied the dynamics of a system of two parametrically driven nonlinear oscillators coupled by a nonlinear interaction.
Several cases have been identified for which analytic solutions for the stationary state are available, while dynamic stability can be numerically assessed.
As a most interesting result, in the symmetric case where the oscillators have equal parameters and are subject to equal driving forces, stationary solutions that spontaneously break the symmetry are found, leading to intriguing multistability phenomena. 

 {The transition from the symmetric solution pair, characterized by equal amplitudes, to the broken symmetry solutions shown in Figs.~\ref{fig:3} and \ref{fig:4}, and illustrated in the phase diagram of Fig.~\ref{fig:5}, corresponds to the branching in the four-dimensional space of the system, shifting from a single amplitude solution to bistable amplitude solutions.}

This behaviour is specific of non-linear systems.
Indeed, in linear symmetric systems under symmetric pumping, the steady state is necessarily symmetric. 
On the other hand, in nonlinear systems, the occurrence of multistability in oscillatory motion allows for the possibility of non-symmetric solutions,
resulting from the emergence of pitchfork bifurcations. 
The set of nonlinear equations governing the complex amplitudes of the system, typically polynomial equations of degree greater than one, can, in fact, yield multiple solutions. 
As parameters change, the system may transition from a single solution to a bifurcation or branching point, where multiple solutions emerge.
The broken symmetry in our system is a manifestation of multistability, with the distinctive feature that branching occurs in a four-dimensional space involving two quadratures for each of the two resonators.

The natural next step will be to investigate the fluctuation properties around the critical points for this multistability phenomenon and the dependence of the switching rate between different solutions on the distance from the critical point.

Another interesting perspective is given by the recent advances in the field of superconducting circuits \cite{Wallraf_RevModPhys.93.025005}, where these types of nonlinear oscillators can be implemented by using various Josephson junction-based technologies \cite{Dykman-book:2012,Hu_PhysRevA.84.012329}.
In this framework, the system can reach a regime where quantum effects become relevant. The pitchfork bifurcation leading to symmetry breaking in the steady-state solutions can eventually be described by a spontaneous symmetry-breaking dissipative phase transition in the quantum regime where the switching process is triggered by quantum fluctuations.
Our findings can thus provide an interesting starting point to explore new types of dissipative phase transition in which the broken symmetry phase is multistable \cite{Savona_PhysRevA.96.033826, Savona_PhysRevA.97.052129, Savona_PhysRevA.101.033839, Domokos_PhysRevX.7.011012,Minganti2023dissipativephase, beaulieu2023observation}.

%
%
%
%
\ack 
%
The work was supported by the Deutsche Forschungsgemeinschaft (DFG, German Research Foundation) 
through Project-ID 25217212 - SFB 1432.
G.R. and I.C. acknowledge
financial support by the PE0000023-NQSTI project by the Italian Ministry of University and Research, co-funded by the
European Union - NextGeneration EU, and 
from Provincia Autonoma di Trento (PAT), partly via the Q@TN initiative.
D.D.B. acknowledges funding from the European Union - NextGeneration EU, "Integrated infrastructure initiative in Photonic and Quantum Sciences" - I-PHOQS [IR0000016, ID D2B8D520, CUP B53C22001750006].
%

\appendix
%
%
%
%
%
%
\section{The model Hamiltonian}
\label{app:model}  
In this appendix we derive Eqs.~(\ref{eq:Main-1}),(\ref{eq:Main-2}). 
We start from the  model Hamiltonian  that describes 
two nonlinear resonators of frequency $\omega_n$ with 
$n=1,2$ and Kerr-Duffing parameter $\kappa_n$
on which one applies one or two parametric external drives 
at frequency $2\omega_d$ and of strength $\mu_n$.

The two modes also have a dispersive interaction of coupling strength $\lambda$.
The Hamiltonian of the system reads
%
%
%
%
%
%
%
%
\begin{equation}
H 
= \sum_{n=1}^{2} H_n
+ \frac{\lambda}{2} q_1^2  q_2^2  
\end{equation}
with
%
%
%
%
%
\begin{equation}
H_n
=
\frac{p_n^2}{2} + \left( \frac{\omega_n^2}{2}   +  \mu_n \cos\left( 2 \omega_d t \right) \right) q_n^2  
+
\frac{\kappa_n}{4} q_n^4
\end{equation}
%
%
%
%
%
%
with $q_n$ and $p_n$ the conjugated variables
(i.e. position and  momentum) of the two modes.
The dynamical equations are given by

%
%
%
%
\begin{equation}
\frac{\partial q_n }{\partial t } = \frac{\partial H}{\partial p_n}  \, , \qquad 
\frac{\partial p_n }{\partial t } = -\frac{\partial H}{\partial q_n} - 2 \Gamma_n \frac{\partial q_n}{\partial t } 
\end{equation}
%
%
%
%
%
%
in which we added a damping force acting on both resonators with damping coefficient $\Gamma_n$.
We apply a canonical transformation
%
%
%
%
%
%
%
%
\begin{equation}
q_n(t) =  \frac{1}{2} u_n(t) e^{i\omega_d t } + \mbox{c.c.} \,, \,\,\,  
p_n(t) =  \frac{i \omega_d}{2} u_n(t) e^{i\omega_d t } - \mbox{c.c.} 
\end{equation}
%
%
%
%
%
%
where $ \mbox{c.c.} $ means complex conjugate.
Here $u_n(t)=X_n(t)-i Y_n(t)$ is the complex amplitude whose components represent 
the two quadratures of the driven motion in  the rotating frame, with $X_n(t)$ the component in phase with the drive 
and $Y_n(t)$ the component out of phase, namely 
$q_n(t) = X_n(t) \cos(\omega_d t) +Y_n(t) \sin(\omega_d t) $.

Using the rotating wave approximation (RWA) with $\omega_d \gg |\omega_d-\omega_n|$, the fast oscillating components of the motion in the rotating frame can be neglected, and the following time-independent equation for the quadratures is obtained
%
%
%
%
%
%
%
%
\begin{equation}
\frac{\partial X_n }{\partial t } = \frac{\partial H_\mathrm{RWA}}{\partial Y_n} 
- \Gamma_n X_n  \, , \quad 
\frac{\partial Y_n }{\partial t }
= -\frac{\partial H_\mathrm{RWA}}{\partial X_n} - \Gamma_n Y_n  \, .
\end{equation}
%
%
%
%
%
%
%
The conservative dynamic is described by the effective Hamiltonian  $H_\mathrm{RWA}$.
Before giving the explicit form of $H_\mathrm{RWA}$, it is useful to scale the quadrature according to the following way
%
%
%
%
%
%
\begin{equation}
X_n = \sqrt{\frac{8\omega_d\Gamma_n}{3\kappa_n}} Q_n \,, \qquad
Y_n = \sqrt{\frac{8\omega_d\Gamma_n}{3\kappa_n}} P_n  
\end{equation}
%
%
%
%
%
%
where $Q_n,P_n$ are dimensionless.
Then, the effective Hamiltonian $H_\mathrm{RWA}$ can be cast as 
%
%
%
%
%
%
\begin{equation}
\tilde{H}_\mathrm{RWA}
=
\frac{3 \sqrt{\gamma_1\gamma_2 }} {8\omega_d \Gamma_1 \Gamma_2} H_\mathrm{RWA} = 
\alpha \, \tilde{H}_1 + \frac{1}{\alpha} \, \tilde{H}_2 + g \,  \tilde{H}_{int}
\end{equation}
%
%
%
%
%
with the parameters 
%
%
%
%
%
%
\begin{equation}
g= \frac{\lambda}{3 \sqrt{\kappa_1\kappa_2}} \,, \quad 
\alpha = \frac{\Gamma_1}{\Gamma_2} \sqrt{\frac{\kappa_2}{\kappa_1}}
\end{equation}
%
%
%
%
%
$g$ represents the scaled coupling strength for the interaction, whereas $\alpha$ is the asymmetry parameter. 
The dimensionless Hamiltonians are given by
%
%
%
%
%
%
\begin{eqnarray}
\tilde{H}_{n} & = \frac{1}{4} {\left(  Q_n^2 + P_n^2 \right)}^2 
- \frac{\delta\Omega_n}{2}  {\left(  Q_n^2 + P_n^2 \right)}
+  \frac{\delta\Omega_{T,n}}{2}  {\left(  Q_n^2 - P_n^2 \right)} \, ,
\\
\tilde{H}_{int}  & = 
 {\left(  Q_1^2 + P_1^2 \right)}  {\left(  Q_2^2 + P_2^2 \right)}
 +
 \frac{1}{2}
 {\left(  Q_1^2-  P_1^2 \right)}  {\left(  Q_2^2 - P_2^2 \right)} 
 +
 2 Q_1 Q_2 P_1 P_2  \, , 
 \phantom{AAAA}
\end{eqnarray}
%
%
%
%
%
and the scaled parameters are 
%
%
%
%
%
%
\begin{equation}
\delta\Omega_n = \frac{\omega_d-\omega_n}{\Gamma_n} \,, \qquad 
\delta\Omega_{T,n} = \frac{\mu_n}{2\omega_d \Gamma_n} 
\end{equation}
%
%
%
%
%
with $\delta\Omega_n$ representing the scaled detuning for the resonator $n$ and 
$\delta\Omega_{T,n} $ is  associated with the frequency threshold for the oscillatory motion of the resonator $n$ at frequency $\omega_d$ 
in the limit of vanishing damping and no interaction $g=0$.
Setting $z_n = Q_n -i P_n$ the dynamical equations read
%
%
%
%
%
%
\begin{eqnarray}
\frac{1}{\Gamma_1} \frac{\partial z_1}{\partial t }
&=
\left( -i\delta\Omega_1 + i {|z_1|}^2 - 1 \right) z_1
+ i \delta \Omega_{T,1} z_1^* 
+ i \frac{g}{\alpha} \left(2  {|z_2|}^2 z_1 + z_2^2 z_1^{*} \right) \\
\frac{1}{\Gamma_2} \frac{\partial z_2}{\partial t }
&=
\left( -i\delta\Omega_2 + i {|z_2|}^2 - 1 \right) z_2
+ i \delta \Omega_{T,2} z_2^*
 + i \alpha g  \left(2  {|z_1|}^2 z_2 + z_2^1 z_2^{*} \right) 
\, .
\end{eqnarray}
%
%
%
%
%
In the main text, 
we assumed that the resonators are almost equal and the asymmetry factor is $\alpha \approx 1$.
More precisely, we consider $\Gamma_1 \approx \Gamma_2$  which implies $\kappa_1 \approx \kappa_2$ 
for $\alpha=1$.
This is valid for $\Gamma_{n} \gg |\Gamma_1-\Gamma_2|$ and  $\kappa_{n} \gg |\kappa_1-\kappa_2|$, i.e. the difference between 
the parameters are small corrections.
Therefore, to simplify the notation, we   set $\Gamma_1\approx \Gamma_2=\Gamma $ and $\kappa_1 \approx \kappa_2  = \kappa$. 
By scaling the time as $\tau = \Gamma t$ we obtain 
Eqs.~(\ref{eq:Main-1}),(\ref{eq:Main-2}).
%
%
%
%
%
%
%
%
\section{The noninteracting case $g=0$ and the steady-state   trivial solution $\bar{z}_1=\bar{z}_2=0$}
\label{app:g=0_z=0}  
In this appendix, we recall the solutions for the noninteracting case $g=0$ 
and analyze the behavior of the trivial solutions $\bar{z}_1=\bar{z}_2=0$.\\
For the uncoupled case $g=0$, we have a single parametrically driven nonlinear resonator. 
Within the RWA and in the limit of vanishing damping, the  equation 
for the stationary solution for the first resonator reads
\begin{equation}
 0 = \left( {|z_1|}^2  - \delta\Omega_1 \right) z_1+  \left( \delta\Omega_{T,1} \right) z_1^{*} 
\label{eq:z1_0}
\end{equation}
and the 
 non-trivial solutions for $z_1$ are simply
\begin{equation}
\bar{z}^{(0)}_{1,+} = \pm i \sqrt{\delta\Omega_1 + \delta\Omega_{T,1} } \,, \,\,\,\,
\bar{z}^{(0)}_{1,-} = \pm  \sqrt{ \delta\Omega_1 - \delta\Omega_{T,1} } \,.
\end{equation}
The solution $\bar{z}^{(0)}_{1,+} $ corresponds to the 
  resonator state when it starts to oscillate at frequency $\omega_d$ above the detuning threshold $ \delta\Omega> - \delta\Omega_{T,1}$ 
and it is completely out of phase with respect to the drive.
By analyzing the dynamics of the harmonic fluctuations around the stationary point, this solution is always dynamically stable above the threshold.
The solution $\bar{z}^{(0)}_{1,-} $ appears above the detuning threshold $ \delta\Omega>  \delta\Omega_{T,1}$ 
and it is in phase respect to the drive: 
This solution is always dynamically unstable above the corresponding threshold. 
The trivial solution $\bar{z}^{(0)}_{1}=0$ is unstable in the detuning range $|\delta \Omega | < \delta\Omega_{T,1}$.

 {
For finite damping the scaled equation reads 
\begin{equation}
\left( {|z_1|}^2  - \delta\Omega_1   + i \right) z_1
= - \delta\Omega_{T,1} z_1^* 
\, ,
\label{eq:z1_0_v2}
\end{equation}
and taking the square modulus of the left and right side of 
Eq.~(\ref{eq:z1_0_v2}), assuming a nontrivial solution $|z_1| \neq 0$, we have ${\left( {|z_1|}^2  - \delta\Omega_1   \right)}^2 =  \delta\Omega_{T,1}^2-1$.
This implies that
}
the effect of finite damping $\Gamma \neq 0$ is a shift of the detuning threshold $\delta\Omega_{T,1} \rightarrow {( \delta\Omega_{T,1}^2-1)}^{1/2}$, 
which reads, using the bare parameters, as the change from  
$\mu_1/(2\omega_d) \rightarrow   {[  {( \mu_1/(2\omega_d)  )}^2 - \Gamma^2   ]}^{1/2}$.

When we switch on the interaction, we discuss the stationary pair solutions.
In this case, the trivial solutions $\left( \bar{z}_{1} =0 , \bar{z}_{2}=0 \right) $ is still a stationary solution which 
remains unstable for $| \delta\Omega_i| < \delta\Omega_{T,i}$ as for the case $g=0$ 
since the equations for the fluctuations around the stationary points are  decoupled and uncorrelated 
as for the noninteracting case $g=0$. 
%
%
%
%
%
%
\section{Solutions with $\bar{z}_2=0$}
\label{app:z2=0}
In this appendix, we discuss the behavior of the solutions of the type $(\bar{z}_1\neq 0, \bar{z}_2=0 )$, namely 
when one of the two resonators does not oscillate.\\
The equations for the fluctuations of the two resonators are decoupled, and in particular, the dynamical equations for the fluctuations 
of the resonator with $\bar{z}_1\neq 0$  remain unchanged respect to the noninteracting case $g=0$.

However, the two resonators are correlated as 
the finite value of the amplitude of the first resonator $\bar{z}_1$ affects the fluctuations $\delta z_2$ of the second one with  $\bar{z}_2=0$.
Therefore, the stability range can differ from that of the noninteracting case.
As a consequence, the stationary pair solution, which is stable for the case $g=0$ 
$\left( \bar{z}_{1,+} = \pm i \sqrt{\delta\Omega + \delta\Omega_{T,1} } , \bar{z}_{2}=0 \right) $
can become unstable.
For example, in the regime  $\omega_1=\omega_2$, $\delta\Omega_{T,2} < g \delta\Omega_{T,1}$ and $g<1/2$, 
the zero amplitude solution with the second resonator $\bar{z}_2=0$ 
becomes unstable in the detuning range
 $\delta\Omega_{T,2} < \delta\Omega <  \left( 2g \delta\Omega_{T,1} -\delta\Omega_{T,2}  \right) / (1-2g) $, well beyond 
the range of the noninteracting case $|\delta\Omega| < \delta\Omega_{T,2}$.
%
%
%
%
%
%
\section{Analytic solutions for $\delta \Omega_{T,1}>0$ and  $\delta \Omega_{T,2}=0$ }
\label{app:single-drive}
In this appendix, we analyze in more detail the pair solution $\bar{z}_1\neq0$  $\bar{z}_2\neq0$ when 
only one resonator is parametrically driven, e.g. $\delta\Omega_{T,1}> 0$ and $\delta\Omega_{T,2}= 0$.
Simple analytical solutions are possible in this case.
To simplify the notation we consider degenerate resonators with $\omega_1 \approx \omega_2$, namely $|\omega_1 - \omega_2| \ll \Gamma$.
If we consider one of the two equations for vanishing parametric drive for the second resonator, we have a simple equation 
\begin{equation}
0
=
\left( {|z_1|}^2  - \delta\Omega  + 2 g  {|z_1|}^2  \right) z_2 +  \left(  z_1^2 \right) z_2^{*} 
\label{eq:z2_z1} \, .
\end{equation}
Comparing Eq.~(\ref{eq:z2_z1})  with Eq.~(\ref{eq:z1_0}) 
we see  that the dispersive interaction has two effects: (i) the amplitude of the first resonator $z_1$ acts as a parametric drive on the second resonator, (ii)
 the amplitude of the first resonator $z_1$ leads to a dispersive frequency shift of the frequency of the second resonator.
This equation can be solved for $z_2$ as function of $z_1$ 
\begin{eqnarray}
z_{2,+}(z_1) &= i \sqrt{\delta\Omega - g  {|z_1|}^2} e^{i \theta_1 } \,, \,\,  \mbox{for} \,\, \delta\Omega - g  {|z_1|}^2 > 0 \\
z_{2,-}(z_1)  &=  \sqrt{\delta\Omega - 3 g  {|z_1|}^2} e^{i \theta_1 } \,, \,\, \mbox{for} \,\, \delta\Omega - 3 g  {|z_1|}^2 > 0 
\end{eqnarray}
with $z_1 = |z_1| e^{i \theta_1 }$.
As a consequence, these two solutions correspond to the two branches of the single parametrically driven Duffing  that we have discussed above, with the difference 
that these two branches do not start symmetrically with respect to zero detuning, but they are centered at finite detuning $2 g  {|z_1|}^2$. 
Then we insert the two solutions $z_{2,\pm}(z_1)$ in the equation $F_1\left(z_1,z_{2,\pm}(z_1)\right) =0 $, in the limit of vanishing damping, such that we obtain 
a close equation for $z_1$.
According to the value of coupling strength $g$ one obtains a zoo of different pair solutions. 
The non-trivial analytic solutions are reported in the table \ref{Tab:1}.

Notice that some solutions exist only in a given range of the coupling strength $g$. In particular different solutions 
appear in the range $0<g<1/3$, $1/3<g<1$ and $g>1$.
The second column of the  table \ref{Tab:1} gives the frequency detuning threshold for the validity of the solutions, whereas the third column refers
to the frequency detuning at which a discontinuity occurs.
When the detuning threshold  coincides with the detuning of the discontinuity, one of the two solutions is characterized by a jump 
from zero to the state of finite amplitude solution. 
When the detuning threshold   is determined by the parametric drive $\delta\Omega_{T,1}$, the solutions are continuous and  only the first derivative 
of the solution with respect to the detuning has a discontinuity.

Table \ref{Tab:1} only reports stationary,  possible solutions.
However, as second step, we must analyze the dynamical stability of these 
stationary solutions.
In general, the region of stability does not coincide with the detuning threshold at which the stationary solution appears, and we must hence compute the stability ranges numerically.

In Fig.~\ref{fig:1}, we reported an example to illustrate the typical behavior of the system with the discontinuity in the first derivative 
of $\bar{z}_1$ as a function of detuning and the parametric-like threshold for $\bar{z}_2$.

\begin{table}[tbhp]
\begin{tabular}{ | p{1.05cm} | p{2.35cm} | p{3.15cm} | p{9cm} | p{1.2cm}|}
\hline \cline{1-5}
 Cond. & Range & Discontinuity & $\bar{z}_1$ &  $\bar{z}_2$ \\
\hline
\cline{1-5}
\hline
$g<1$	& $\delta\Omega>- \delta\Omega_{T,1}$			& $ \delta\Omega_{c,1}  = \frac{g}{1-g} \delta\Omega_{T,1}$
		&  $\pm i \sqrt{\delta\Omega+ \delta\Omega_{T,1} - \theta\left( \delta\Omega - \delta\Omega_{c,1} \right) \frac{g}{1+g} \left( \delta\Omega - \delta\Omega_{c,1}  \right) }$
		&  $\mp  |\bar{z}_{2,+}| $ \\
\hline
$g<1$	& $\delta\Omega> \delta\Omega_{c,2}$			& $ \delta\Omega_{c,2}  = \frac{1}{1-g} \delta\Omega_{T,1}$
		&  $\pm  \sqrt{\frac{1}{1+g} \left( \delta\Omega - \delta\Omega_{c,2}  \right) }$
		&  $\pm i   |\bar{z}_{2,+}|$ \\
\hline
$1<g$	& $\delta\Omega> \delta\Omega_{c,1}^*$			& $ \delta\Omega_{c,1}^*  = \frac{1}{g-1} \delta\Omega_{T,1}$
		&  $\pm i  \sqrt{\frac{1}{1+g} \left( \delta\Omega - \delta\Omega_{c,1}^*  \right) }$
		&  $\mp    |\bar{z}_{2,+}|$ \\		
\hline
$1<g$	& $\delta\Omega> \delta\Omega_{T,1}$			& $ \delta\Omega_{c,2}^*  = \frac{g}{g-1} \delta\Omega_{T,1}$
		&  $\pm  \sqrt{\delta\Omega- \delta\Omega_{T,1} - \theta\left( \delta\Omega - \delta\Omega_{c,2}^* \right) \frac{g}{1+g} \left( \delta\Omega - \delta\Omega_{c,2}^*  \right) }$
		&  $\pm i    |\bar{z}_{2,+}|$ \\		
\hline
\cline{1-5}
\hline
$g<\frac{1}{3}$	& $\delta\Omega>- \delta\Omega_{T,1}$			& $ \delta\Omega_{c,3}  = \frac{3g}{1-3g} \delta\Omega_{T,1}$
			&  $\pm i \sqrt{\delta\Omega+ \delta\Omega_{T,1} - \theta\left( \delta\Omega - \delta\Omega_{c,3} \right) \frac{3g}{1+3g} \left( \delta\Omega - \delta\Omega_{c,3}  \right) }$
			&  $\pm i  |\bar{z}_{2,-}| $ \\
\hline
$g<\frac{1}{3}$	& $\delta\Omega> \delta\Omega_{c,4} $			& $ \delta\Omega_{c,4}  = \frac{1}{1-3g} \delta\Omega_{T,1}$
			&  $\pm  \sqrt{\frac{1}{1+3g} \left( \delta\Omega - \delta\Omega_{c,4}  \right) }$
			&  $\pm |\bar{z}_{2,-}|  $ \\
\hline
$\frac{1}{3}<g$	& $\delta\Omega> \delta\Omega_{c,3}^* $			& $ \delta\Omega_{c,3}^*  = \frac{1}{3g-1} \delta\Omega_{T,1}$
			&  $\pm i  \sqrt{\frac{1}{1+3g} \left( \delta\Omega - \delta\Omega_{c,3}^*  \right) }$
			&  $\pm i  |\bar{z}_{2,-}|  $ \\
\hline
$\frac{1}{3}<g$	& $\delta\Omega> \delta\Omega_{T,1} $			& $ \delta\Omega_{c,4}^*  = \frac{3g}{3g-1} \delta\Omega_{T,1}$
			&  $\pm  \sqrt{\delta\Omega - \delta\Omega_{T,1} - \theta\left( \delta\Omega - \delta\Omega_{c,4}^* \right) \frac{3g}{1+3g} \left( \delta\Omega - \delta\Omega_{c,4}^*  \right) }$
			&  $\pm   |\bar{z}_{2,-}|  $ \\			
\hline
\end{tabular}
\caption{Non-trivial analytic solutions of the stationary state for a single driven mode, for the symmetric case, and in the limit of 
vanishing damping. The solution for $\bar{z}_2$  are  $|\bar{z}_{2,+}| = \sqrt{\delta\Omega - g  {|\bar{z}_1|}^2} $ and $|\bar{z}_{2,-}| =  \sqrt{\delta\Omega - 3 g  {|\bar{z}_{1}|}^2} $ (see text).} 
\label{Tab:1}
\end{table}

%
%
%
\section*{References}
\bibliographystyle{unsrt}

\bibliography{bibliography}
%
%
%

\end{document}